\documentstyle[epsfig]{aipproc}

\begin{document}
\title{Final results from the NMC}

\author{Eva-Maria Kabu{\ss}\thanks{supported by the BMBF}}
\address{Inst. f\"ur Kernphysik, University of Mainz\\ Becherweg 45,
55099 Mainz}
\smallskip
\address{for the NEW MUON COLLABORATION}
\maketitle

\begin{abstract}
A summary of the results on deep inelastic muon nucleon and muon
nucleus scattering experiments performed by the NMC collaboration is
presented.
\end{abstract}

The New Muon Collaboration (NMC) has measured deep inelastic muon scattering
at the M2 muon beam line of the CERN SPS 
at incident energies between 90 and 280 GeV. The data taken between 1986 and 
1989 cover three main topics:
\begin{itemize}
\item A detailed study of the structure of free nucleons by measuring
$F_2^{\rm p}$, $F_2^{\rm d}$, $R$, $R^{\rm d}-R^{\rm p}$ and 
$F_2^{\rm d}/F_2^{\rm p}$ using hydrogen and deuterium
targets.
\item A comparison of bound and free nucleons to study nuclear effects with a 
series of targets ranging from deuterium to lead.
\item A study of the $Q^2$ dependence 
of nuclear effects using high 
luminosity measurements with thick carbon and tin targets.
\end{itemize}
All measurements were done simultaneously with two target materials in the muon
beam which were frequently interchanged with a complementary target set
where the sequence of the materials along the beam line was reversed.
This results in a cancellation of
systematic errors due to spectrometer acceptance and normalisation for the 
determination of cross section ratios and allows  
measurements in kinematic regions where
the detector acceptance was small \cite{emk_nmclongrat}.
In the extraction of the structure functions $F_2^{\rm p}$ and $F_2^{\rm d}$ the
spectrometer acceptance was substantially different for the upstream and
the downstream targets, thus giving two separate measurements for each 
material allowing a good control of systematic errors due to reconstruction and
detector acceptance \cite{emk_nmcf292}.

Measurements with hydrogen and deuterium were done to determine the dependence
on the Bjorken scaling variable $x$ and the four momentum squared  $Q^2$
of the structure functions $F_2^{\rm p}$ and $F_2^{\rm d}$. 
As data were taken
at four incident energies also $R$, the ratio of the longitudinally 
to transversely polarised virtual photon absorption cross sections, could be
extracted from the small differences in cross section at the same values of 
$x$ and $Q^2$ but at different values of the relative energy transfer $y$, 
i.e. at different beam energies. Due to the use
of a small angle trigger the covered $x$ and $y$ range was 
extended downwards, thus increasing the sensitivity to $R$ \cite{emk_nmc_f296}. 
We chose to 
determine $R(x)$ averaged over $Q^2$ as the statistics of the data 
($2.4 \cdot 10^6$
events) did not allow to determine both its $x$ and $Q^2$ dependence. 

\begin{figure}
\begin{center}
\begin{tabular}[h]{ll}
\epsfig{file=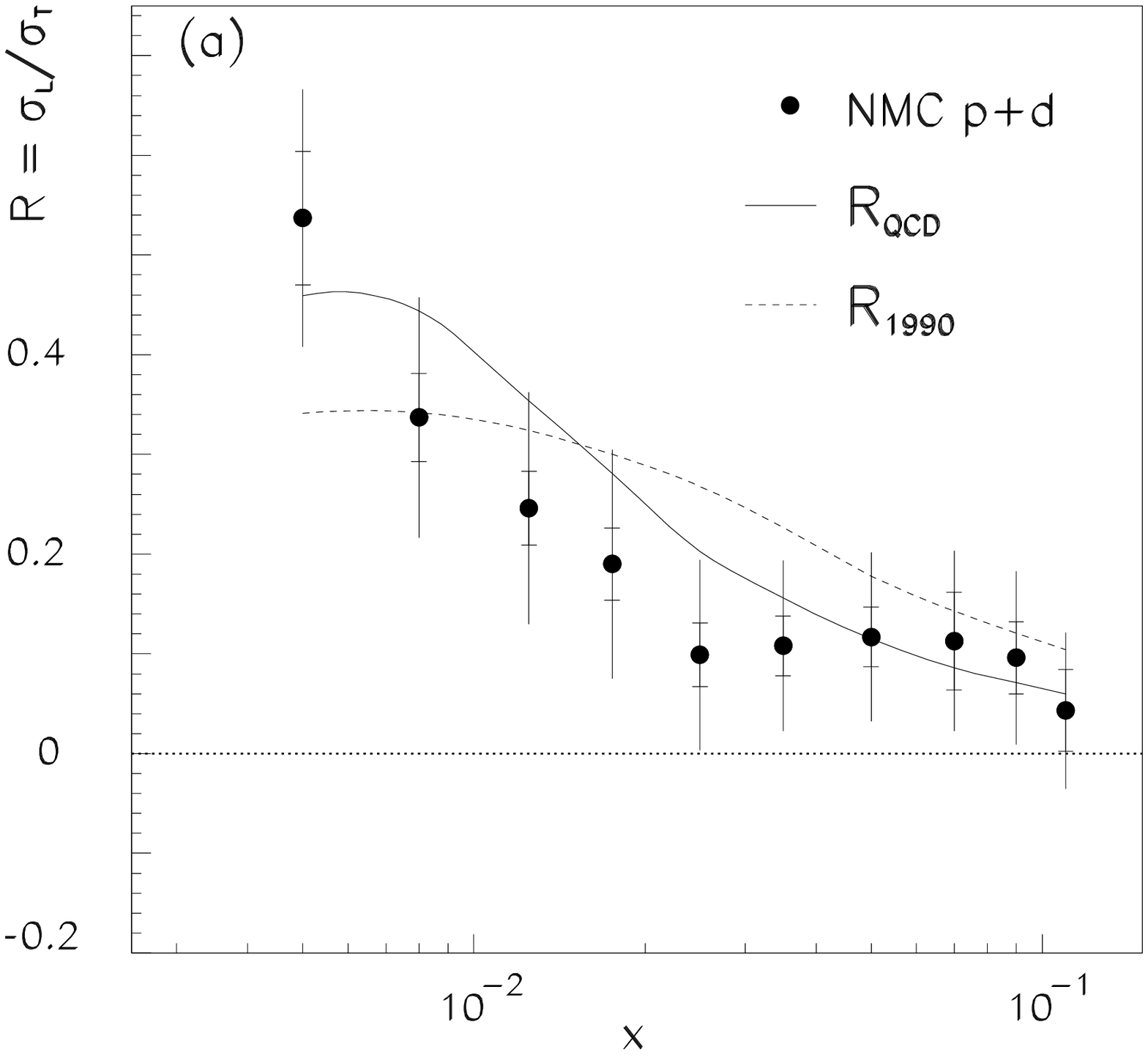,width=.475\textwidth} &
\epsfig{file=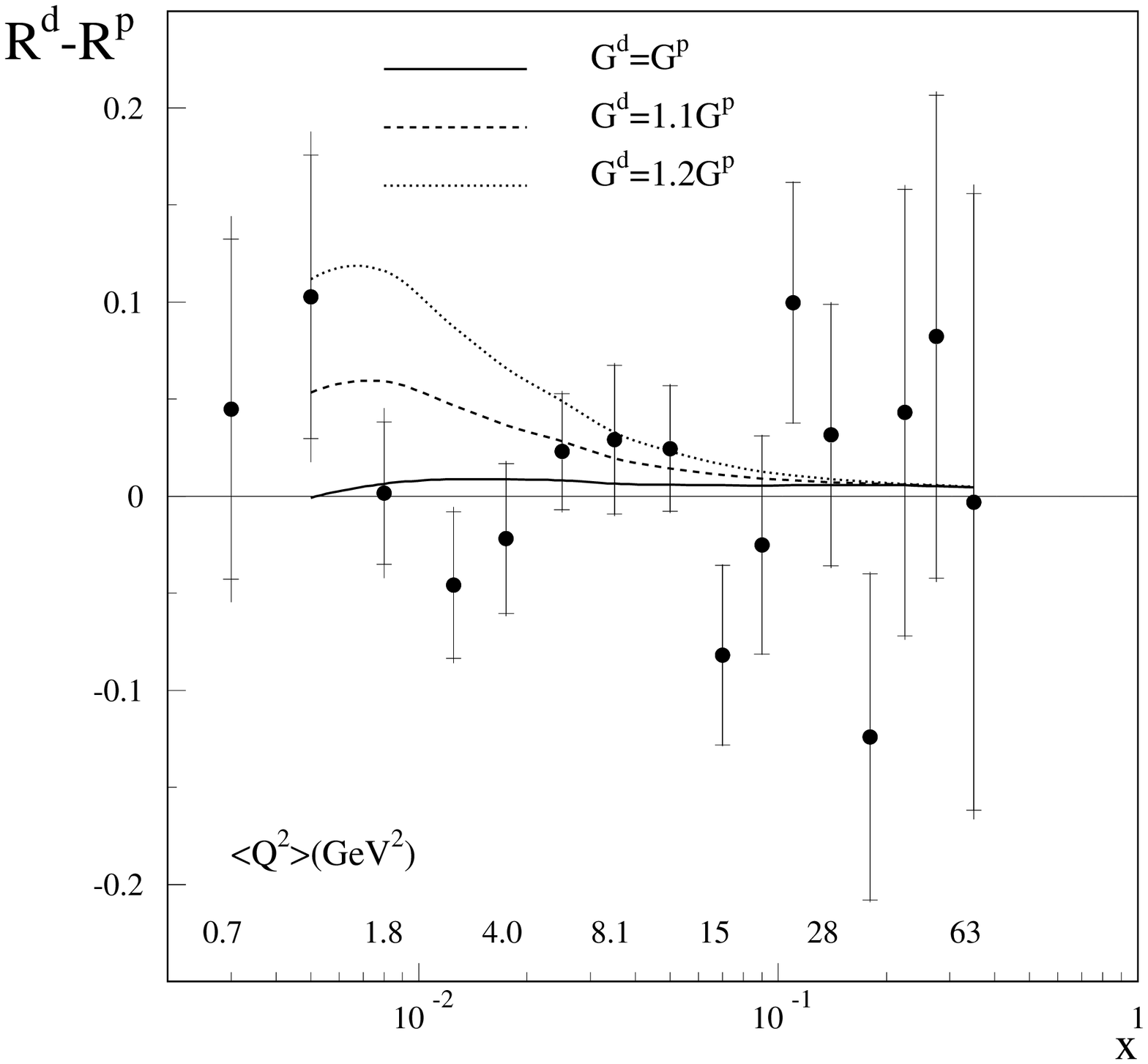,width=.475\textwidth}
\end{tabular}
\end{center}
\caption{Left: Results for $R(x)$ averaged for the proton and the deuteron.
The curves are discussed in the text. Right: Results for 
$R^{\rm d}-R^{\rm p}(x)$ 
compared to predictions from perturbative QCD (see text). The inner error bars
correspond to statistical errors and the full ones to the quadratic sum of 
statistical and systematic errors.}
\label{emk_fig_r} 
\end{figure}

The results for $R$
are shown in fig.\ref{emk_fig_r} (left) in the range $0.002<x<0.12$ 
compared to a QCD 
prediction (solid line) obtained using a QCD analysis of part of the data
\cite{emk_nmc_qcd} and a parametrisation of previous $R$ measurements 
\cite{emk_rslac}.
Within the largely correlated systematic errors good agreement is observed.

\begin{figure}
\begin{center}
\begin{tabular}[h]{ll}
\epsfig{file=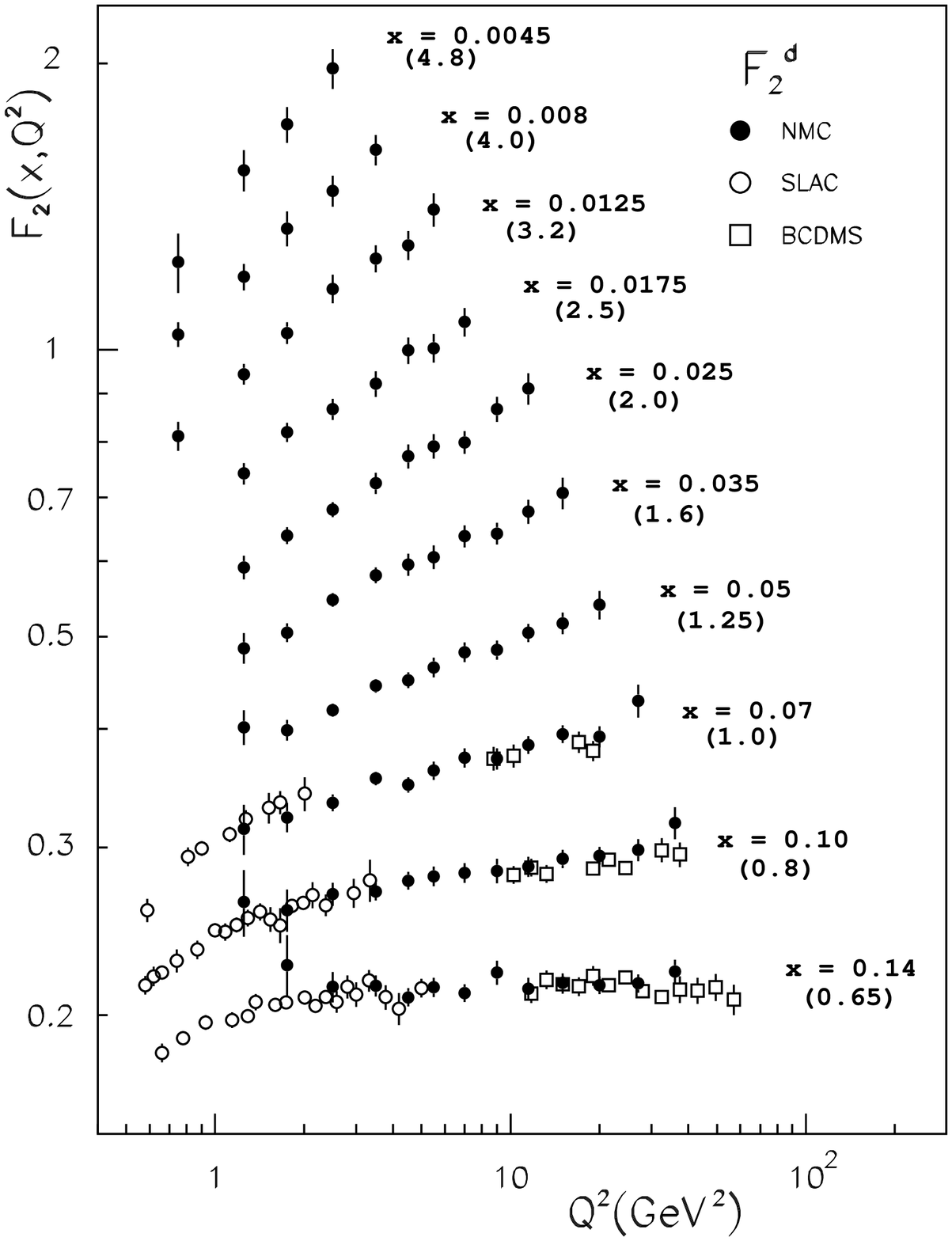,width=.475\textwidth} &
\epsfig{file=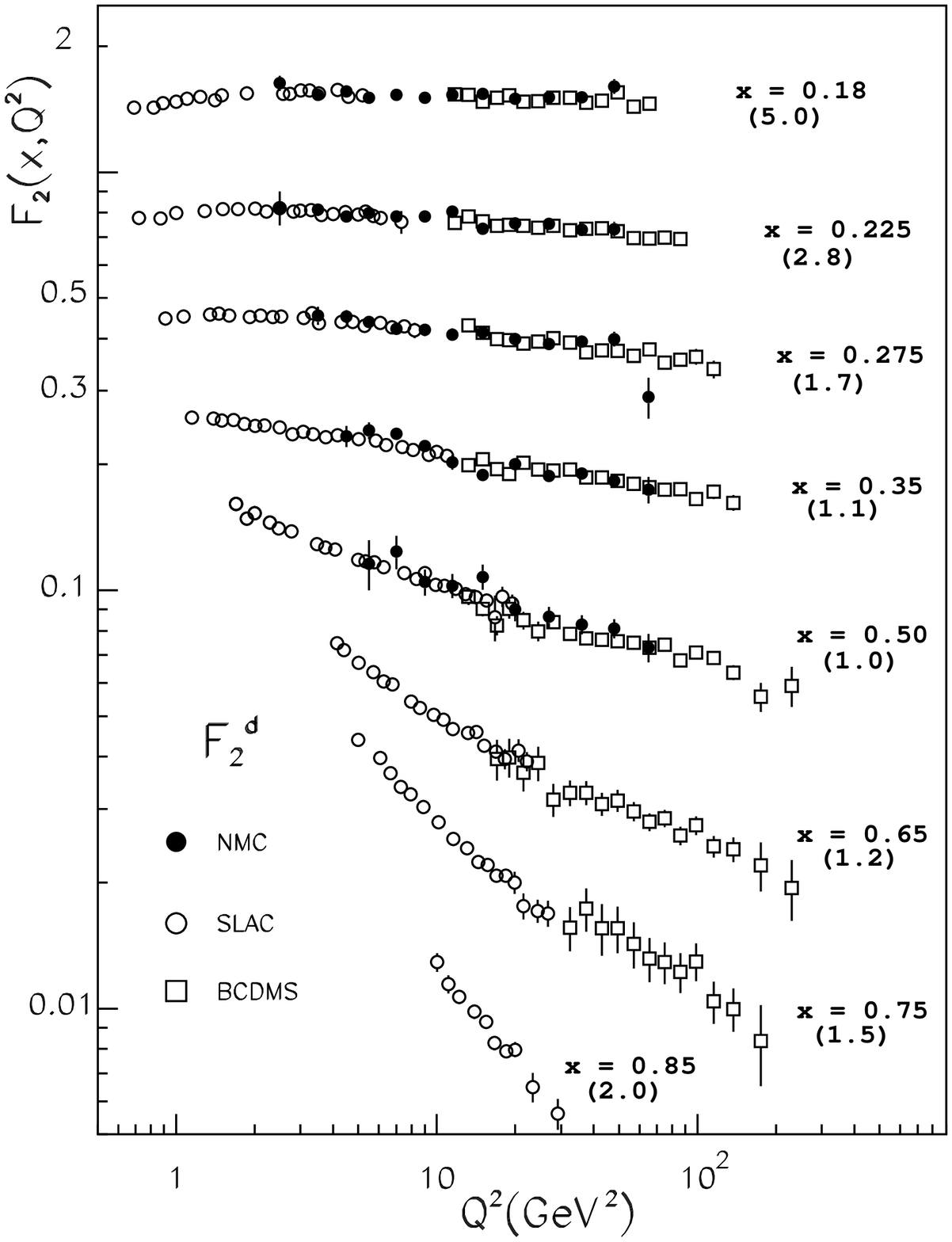,width=.475\textwidth}
\end{tabular}
\end{center}
\caption{NMC results for $F_2^{\rm d}$ compared to SLAC and BCDMS results. The error 
bars represent the total errors, apart from normalisation uncertainties.}
\label{emk_fig_f2} 
\end{figure}

In the extraction of structure functions
the SLAC $R$ parametrisation was used for $x>0.12$.
The results for $F_2^{\rm d}$ covering a kinematic range of 
$0.0045<x<0.6$ and $0.5<Q^2<75$ GeV$^2$ are shown in fig.\ref{emk_fig_f2}.
The additional normalisation error of the NMC data is 2.5\%. The results for
$F_2^{\rm p}$ and $F_2^{\rm d}$ compare well with previous measurements from 
SLAC \cite{emk_slac_f2} and BCDMS \cite{emk_bcdms_f2}
(see figure) and E665 \cite{emk_e665_f2} and extrapolate smoothly to 
the recent H1 and ZEUS data.

From the same data $R^{\rm d}-R^{\rm p}$ and $F_2^{\rm d}/F_2^{\rm p}$ could 
be determined in an 
enlarged kinematic range due to the use of the complementary targets 
($8.4\cdot 10^6$ events) \cite{emk_nmc_np97}.
The results for $R^{\rm d}-R^{\rm p}$ are shown in fig.\ref{emk_fig_r} (right) 
in the $x$ range
from 0.003 to 0.35. The values of $\Delta R$ are small; this is especially 
significant at small $x$ where $R$ is large. No significant $x$ dependence is
observed. Averaging the measurements one obtains at $Q^2=10$~GeV$^2$ a value
of $R^{\rm d}-R^{\rm p}=0.004 \pm 0.012 ({\rm stat.}) \pm 0.011({\rm syst.})$ 
compatible with zero.
The results are compared to NLO perturbative QCD computations of 
$\Delta R$ using 
various assumptions for the difference of the gluon distribution in the
proton and the deuteron showing that the data set a limit of about 10\% to this
difference.

\begin{figure}
\begin{center}
\begin{tabular}[h]{ll}
\epsfig{file=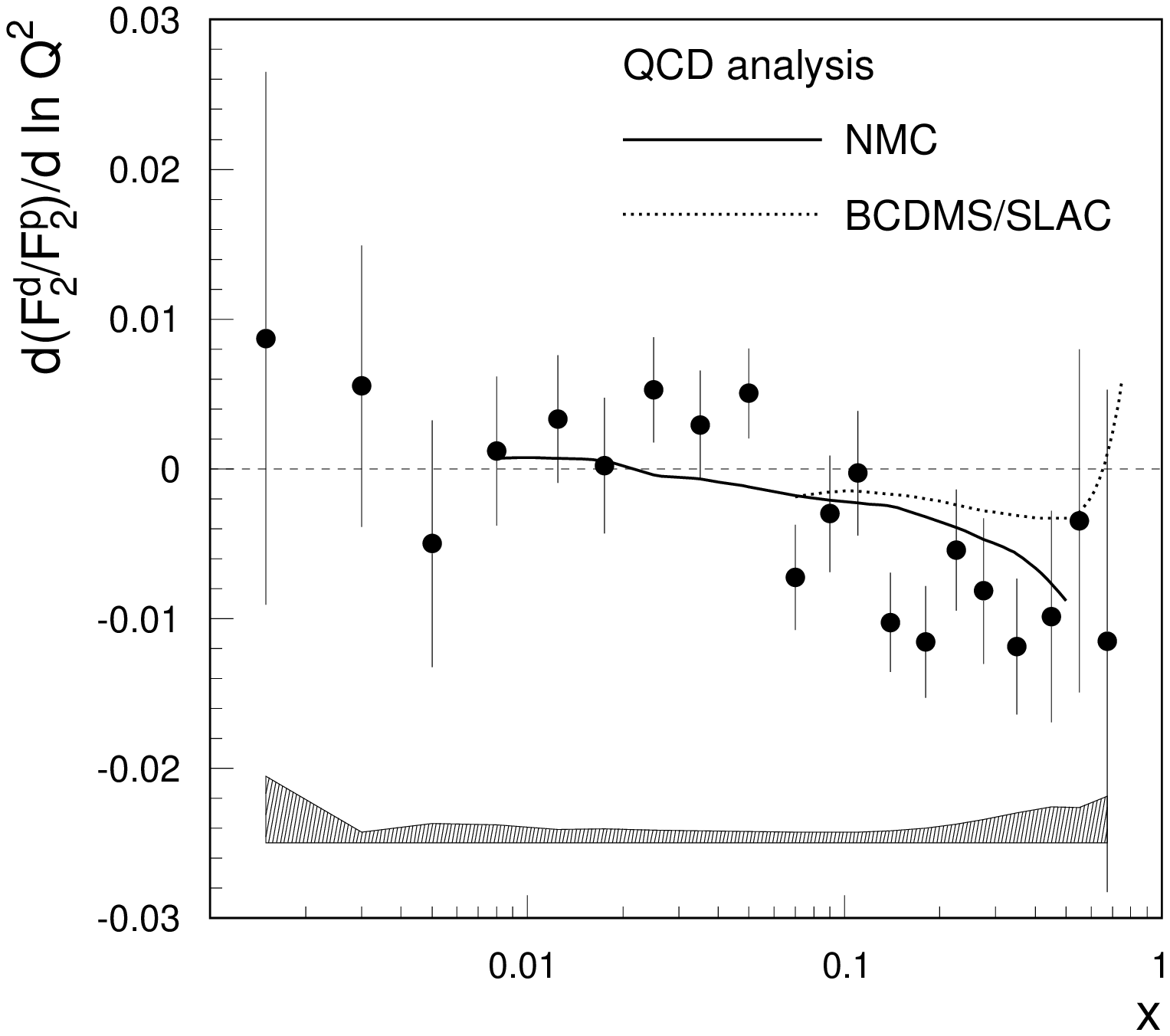,width=.475\textwidth} &
\epsfig{file=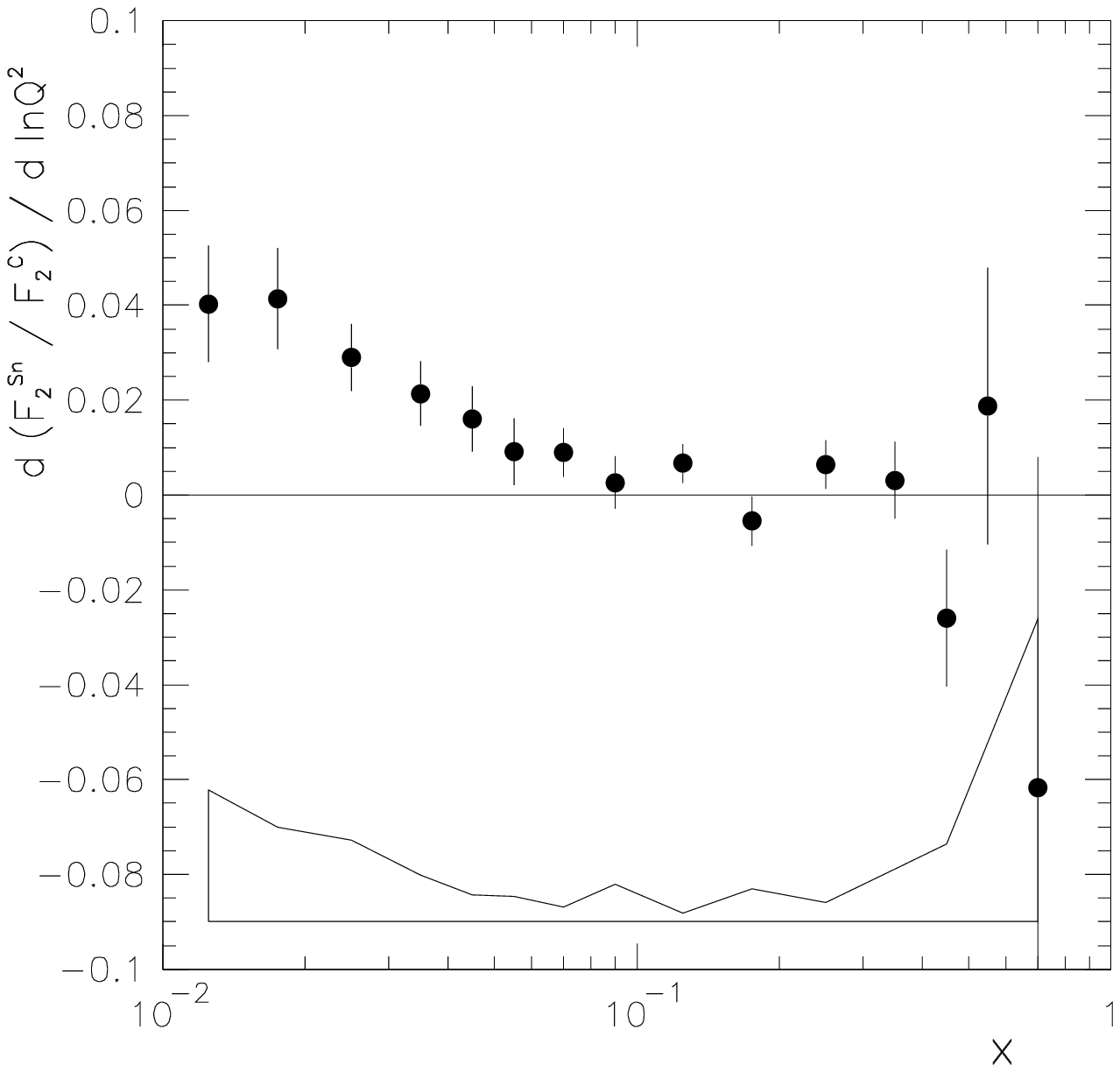,width=.455\textwidth}
\end{tabular}
\end{center}
\caption{Left: The logarithmic slopes 
${\rm d}(F_2^{\rm d}/F_2^{\rm p})/{\rm d}\ln Q^2$, curves see text.
Right: The logarithmic slopes ${\rm d}(F_2^{\rm SN}/F_2^{\rm C})/{\rm d}\ln Q^2$.
The error bars represent the statistical uncertainty. The bands show the sizes
of the systematic uncertainties. }
\label{emk_fig_slopes} 
\end{figure}

The results for the structure function ratio $F_2^{\rm d}/F_2^{\rm p}$ cover the 
$x$ range 
from 0.001 to 0.8 and the $Q^2$ range from 0.1 to 145~GeV$^2$ with high accuracy,
thus allowing to investigate the $Q^2$ dependence in a large $x$ range. Fitting
a linear function in $\ln Q^2$ at fixed $x$ to the results yields the 
logarithmic slopes shown in fig.\ref{emk_fig_slopes} (left). They are compared
to NLO QCD calculations based on analyses of the NMC \cite{emk_nmc_qcd} and 
the SLAC/BCDMS \cite{emk_mv_qcd} data.
The measured slopes are consistent with these calculations although there may 
be deviations at $x>0.1$, as was suggested in \cite{emk_nmclongrat}. 

Averaging over $Q^2$ one obtains results for the $x$ dependence of 
$F_2^{\rm d}/F_2^{\rm p}$ with total errors of less than 1\% in a large range 
of $x$. The ratio shows the well known $x$ dependence dropping to about 0.7 at
high $x$ and approaching unity at small $x$ indicating no sizeable shadowing.
From the results for $F_2^{\rm d}/F_2^{\rm p}$ the Gottfried sum 
$S_G=\int (F_2^{\rm p}-F_2^{\rm n}){\rm d}x/x$ can be calculated using a 
parametrisation of $F_2^{\rm d}$. In the measured range of $0.004<x<0.8$ 
one obtains
a contribution of $0.2281\pm 0.0065 ({\rm stat.})$ to $S_G$ at $Q^2=4$~GeV$^2$.
This agrees within statistical errors with the previously published value
\cite{emk_nmc_gott}.

Nuclear effects were investigated studying the dependence on the mass number 
$A$ in the shadowing region (small $x$), 
the enhancement region (at $x \approx 0.1$)
and the EMC effect region (large $x$) by measuring with a series of nuclei.
A clear increase with 
$A$ was observed for all effects  \cite{emk_adep}.

A more detailed study of the ratio $F_2^{\rm Sn}/F_2^{\rm C}$ was performed by 
using a special high luminosity setup with a total target thickness of
600~g/cm$^3$ and an active target calorimeter to trigger on DIS events.
Due to the increased multiple Coulomb scattering tighter kinematic cuts had to 
be applied resulting in a more limited range in $x$ and $Q^2$ and a Monte
Carlo simulation of vertex migration and kinematic smearing was necessary.
Measurements were done at 280, 200 and 120 GeV resulting in 
$8.4 \cdot 10^6$ events \cite{emk_qdep}.

\begin{figure}
\begin{center}
\begin{tabular}[htb]{ll}
\epsfig{file=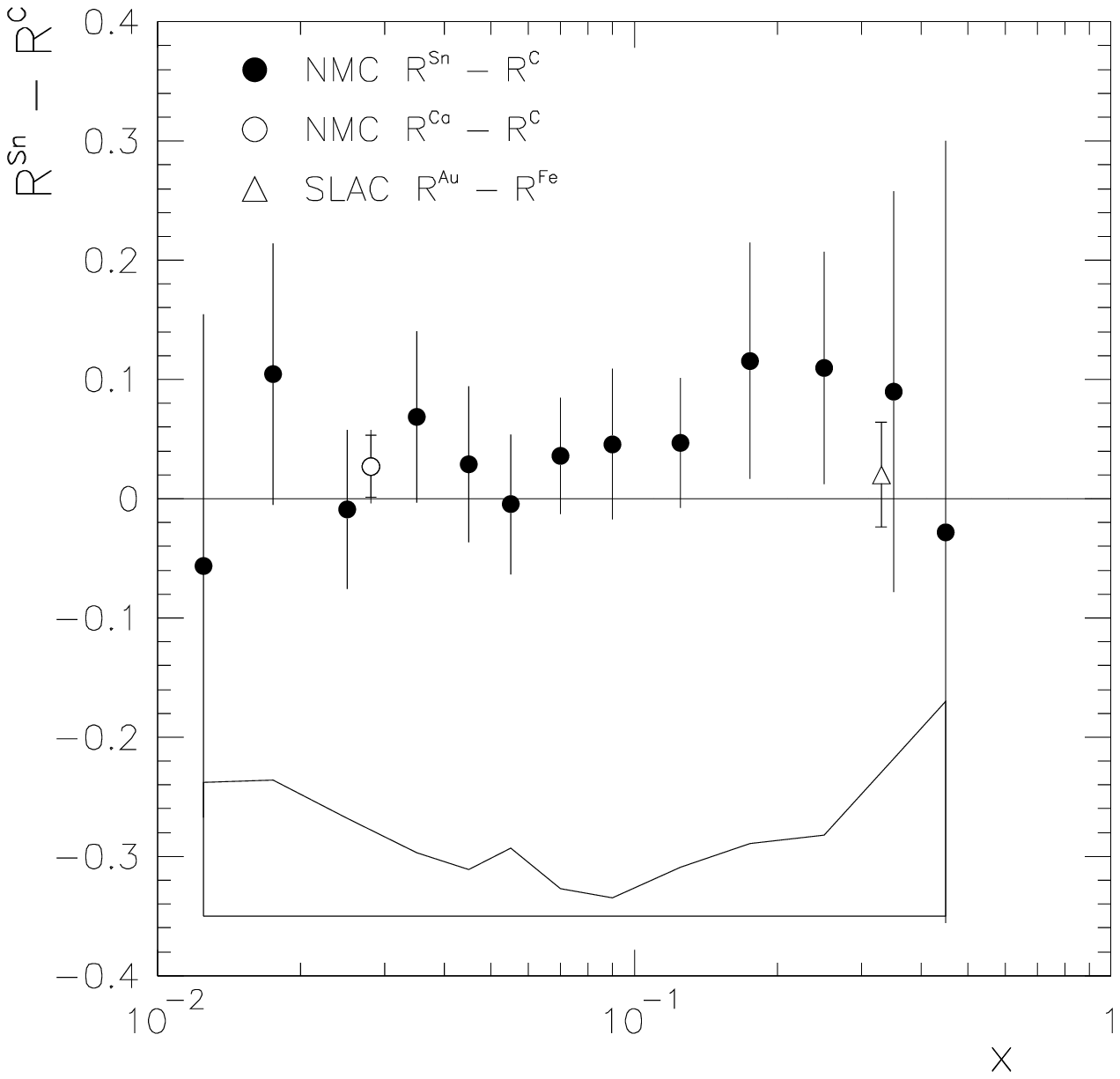,width=.45\textwidth}
\epsfig{file=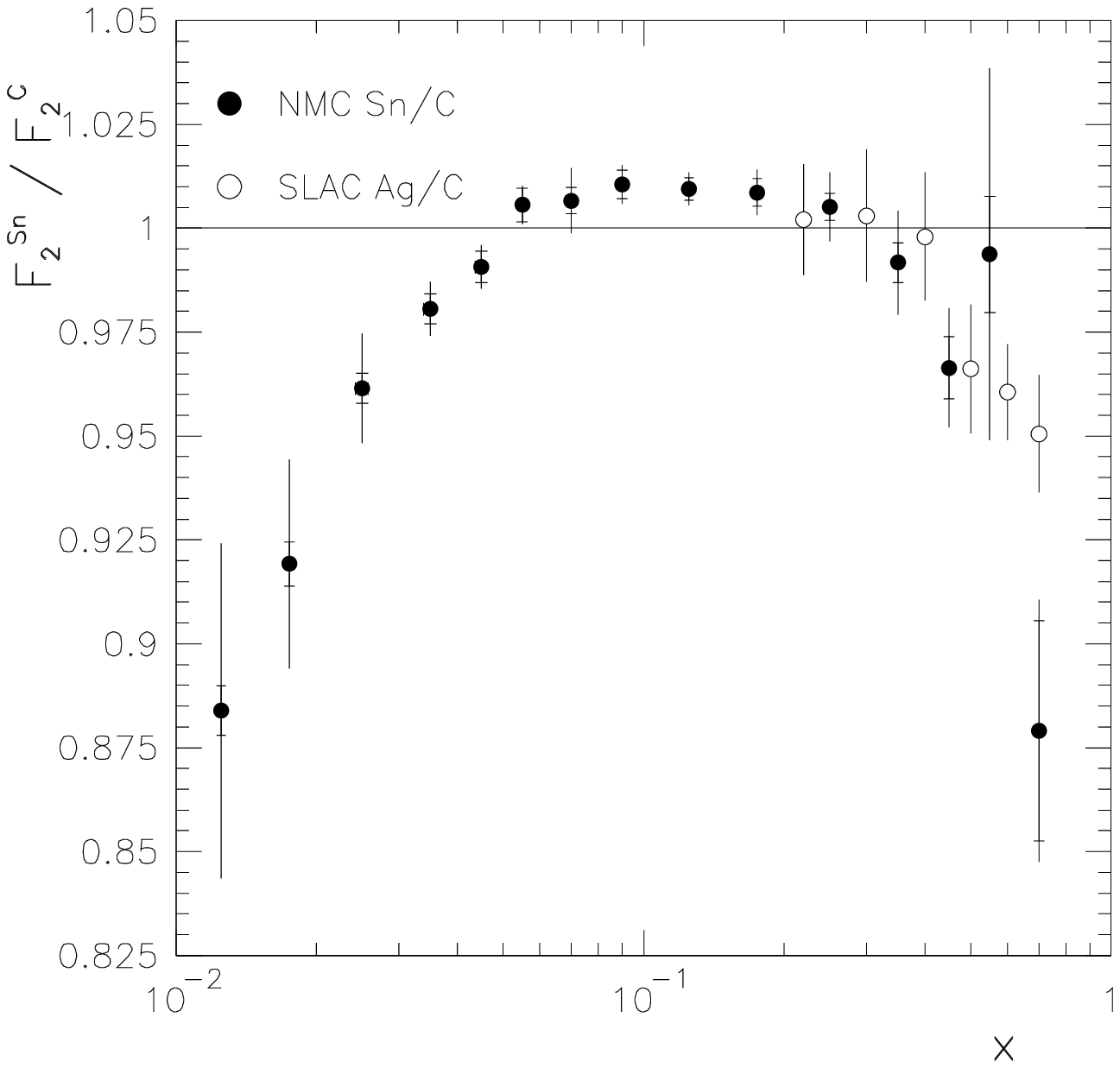,width=.45\textwidth} &
\end{tabular}
\end{center}
\caption{Left:
$R^{Sn}-R^C$ as a function of $x$, averaged over $Q^2$.
Also shown are the NMC result for the average value of
$R^{Ca}-R^C$~\protect\cite{emk_nmc_dr} (open circle) and
the SLAC one for  $R^{Au}-R^{Fe}$~\protect\cite{emk_slac_e140} (triangle).
Right: $F_2^{\rm Sn}/F_2^{\rm C}$ as a function of $x$, averaged over
$Q^2$. The SLAC-E139~\protect\cite{emk_slac_52} ratios
for silver and carbon are also plotted (open points).}
\label{emk_fig_sn}
\end{figure}

The results for $R^{\rm Sn}-R^{\rm C}$ are shown in fig.\ref{emk_fig_sn} (left)
for the $x$ range of 0.01 to 0.5. No $x$ dependence is observed and the 
average value is $0.040 \pm 0.21({\rm stat.}) \pm 0.026 ({\rm syst.})$
at a mean $Q^2$ of 10 GeV$^2$.
The results for the $x$ dependence of $F_2^{\rm Sn}/F_2^{\rm C}$ in the $x$ range
$0.01<x<0.75$ in fig.\ref{emk_fig_sn} (right) confirm the well known $x$ dependence
of nuclear effects in the structure function $F_2$ and give a very precise 
measurement of the small enhancement of about 1\% at $x\approx 0.1$. The $Q^2$
dependence was investigated in the range $1<Q^2<140$~GeV$^2$. Fitting
linear functions in $\ln Q^2$ to the results at fixed $x$ yields the 
logarithmic slopes shown in fig.\ref{emk_fig_slopes} (right), indicating
that the amount of shadowing decreases with $Q^2$. No $Q^2$ dependence is 
observed at higher $x$.

\end{document}